\documentclass[twocolumn,trackchanges]{aastex631}

\newcommand{\bpx}{BPX}

\submitjournal{ApJ}

\shorttitle{Secular Attrition of Classical Bulges by Stellar Bars}
\shortauthors{R.L. McClure et al.}

\usepackage{xcolor}
\usepackage{ulem}   
\definecolor{kcolor}{rgb}{0.54, 0.17, 0.89}

\begin{document}

\title{The Secular Attrition of Classical Bulges by Evolving Stellar Bars}

\author[0000-0001-5928-7155]{Rachel Lee McClure}
\correspondingauthor{Rachel Lee McClure}
\email{rlmcclure@wisc.edu}
\affiliation{University of Wisconsin–Madison \\
Department of Astronomy \\
Madison, WI, 53706, USA}

\author[0000-0002-6851-9613]{Tobias G\'eron}
\affiliation{David A. Dunlap Institute of Astronomy \& Astrophysics\\
University of Toronto\\
Toronto, ON, M5S 3H4, Canada}

\author{Elena D'Onghia}
\affiliation{University of Wisconsin–Madison \\
Department of Astronomy \\
Madison, WI, 53706, USA}

\author[0000-0002-8658-1453]{Angus Beane}
\affiliation{Center for Astrophysics $|$ Harvard \& Smithsonian, 60 Garden Street, Cambridge, MA 02138, USA}

\author{Aaryan Thusoo}
\affiliation{David A. Dunlap Department of Astronomy \& Astrophysics\\
University of Toronto\\
Toronto, ON, M5S 3H4, Canada}

\author[0000-0003-2594-8052]{Kathryne J. Daniel}
\affiliation{Department of Astronomy \& Steward Observatory, University of Arizona,
Tucson, AZ 85721, USA}

\author[0000-0001-5522-5029]{Carrie Filion}
\affiliation{Center for Computational Astrophysics, Flatiron Institute, 162 Fifth Avenue, New York, NY 10010, USA}

\author[0000-0001-9982-0241]{Scott Lucchini}
\affiliation{Center for Astrophysics $|$ Harvard \& Smithsonian, 60 Garden Street, Cambridge, MA 02138, USA}

\begin{abstract}
Classical bulges and stellar bars are common features in disk galaxies and serve as key tracers of galactic evolution.  Angular momentum exchange at bar resonances drives secular morphological changes throughout the disk, including bar slowing and lengthening, and affects the structure of accompanying bulges. In this study, using a suite of N-body simulations, we quantify the secular reconfiguration of classical bulges through resonant trapping by evolving stellar bars. 
We use orbital frequency analysis to identify bar-resonant populations and find that up to 50\% of the initial bulge stars become trapped in 2:1 resonant orbits and adopt disk-like kinematics. This transformation renders much of the classical bulge observationally indistinguishable from the disk. We compare these results with a sample of 210 MaNGA disk galaxies, finding that slow bars-—indicative of older systems-—are preferentially associated with weaker bulges. These results suggest that long-lived bars can significantly reshape classical bulges, potentially explaining their scarcity in the local universe and the low classical bulge fraction found in the Milky Way.

\end{abstract}


\section{Introduction}

Disk galaxies are ubiquitous in our local universe \citep{2010ApJ...723...54K,2010A&A...509A..78D,2014ApJ...787...24B,2014MNRAS.444.1647K} and often host central stellar structures, including stellar bulges and bars. Recent results show that disk galaxies dominate in fraction over their peculiar and elliptical counterparts up to redshifts of $z\sim6$ \citep{2010A&A...509A..78D,2023ApJ...955...94F} and that stellar bars form in these galactic disks as early as $z\sim4.2$ \citep{2023ApJ...945L..10G, 2023ApJ...947...80B, 2023ApJ...948L..13J, 2023ApJ...955...94F, 2024arXiv240906100G, 2024MNRAS.530.1984L, 2025arXiv250501421G}. Bulges are divided into random motion dominated classical bulges formed through hierarchical assembly processes \citep{1982ApJ...256..460K, 1982ApJ...257...75K,1993MNRAS.264..201K, 1997ARA&A..35..637W, 2002NewA....7..155S} and pseudobulges that are rotationally supported \citep{2010ApJ...715L.176K}, while stellar bars emerge later through either internal disk instabilities or external tidal interactions. Angular momentum exchange can build rotating pseudobulge features that include stars that originate in both the stellar classical bulge and disk. Understanding this morphological transition is essential for interpreting how disk galaxies evolve structurally across cosmic time.

Early type disks (S0-Sb) are observed with a central bulge ($B/T\gtrsim0.2$), while late-type disks have minimal bulge-to-total luminosities \citep{2001A&A...367..428A, 2004ARA&A..42..603K, 2006A&A...457...91E, 2009ApJ...696..411W}. Rotating central components of observed bulges, classified as pseudobulges, are observed in 25--50\% of late-type spirals throughout the local universe \citep{2004ARA&A..42..603K, 2008AJ....136..773F}. Pseudobulges (including Boxy/Peanut X-Features) evolve through secular, internal dynamical processes that alter the orbits of disk stars to build a bulge-like central component. This process has been studied in cosmological \citep[e.g.][]{2015MNRAS.454.1886S, 2020MNRAS.494.5936F} and isolated simulations \citep{1982ApJ...256..460K, 2002AJ....124..722Q,2004ARA&A..42..603K,2014ApJ...787L..19n, 2024ApJ...975..120T, 2025MNRAS.537.1475M}. The emergence of pseudobulge occurs primarily through bar destabilization \citep[e.g.][]{1981A&A....96..164C,1991Natur.352..411R, 1994ApJ...425..551M, 2003MNRAS.346..251O, 2004ApJ...613L..29M, 2005MNRAS.364L..18B,2006ApJ...645..209D, 2008IAUS..245...93A} and resonance passage \citep[e.g.][]{2014MNRAS.437.1284Q,2020MNRAS.495.3175S,2024A&A...692A.145Z, 2025MNRAS.537.1475M, 2025arXiv250112308L}. 

Including our host Milky Way disk galaxy, more than 60\% of massive nearby disks host a central stellar bar \citep{2000AJ....119..536E, 2004ApJ...612..191E, 2005MNRAS.364..283E, 2008ApJ...675.1141S, 2009A&A...495..491A, 2009MNRAS.393.1531G, 2010MNRAS.405..783M, 2013ApJ...779..162C,2018MNRAS.473.4731K,2018MNRAS.474.5372E,2023MNRAS.524.3166E}. Stellar bars are resonance features \citep{1972MNRAS.157....1L, 1977A&A....61..477C, 1979MNRAS.187..101L, 1981A&A....99..362S} that grow in prominence and length over time \citep{2002ApJ...569L..83A, 2024A&A...683A.100G}.
Simulations show they can emerge through both external gravitational perturbing events like fly-bys or mergers \citep{1990A&A...230...37G, 2014MNRAS.445.1339L,2016ApJ...826..227L, 2019MNRAS.483.2721P}, and through internal excitation of underlying modes in the differentially rotating disk \citep{1971ApJ...168..343H, 1973ApJ...186..467O, 1987MNRAS.225..653S, 2003MNRAS.341.1179A, 2013MNRAS.429.1949A}. 

The formation of a stellar bar can mark a period in which secular, internal processes begin to dominate the evolution of structure within a galaxy \citep[e.g.][]{2004ARA&A..42..603K}. 
Strong bars power significant responses in non-bar orbits throughout the galactic disk by angular momentum exchange with the stellar halo and stellar bulge \citep{2003MNRAS.341.1179A,2012MNRAS.421..333S,2013MNRAS.430.2039S,2018MNRAS.475.1653K}, the dark matter halo \citep{1984ApJ...282L...5T, 1985MNRAS.213..451W, 1992ApJ...400...80H, 2000ApJ...543..704D, 2002MNRAS.330...35A, 2002ApJ...569L..83A, 2007MNRAS.375..460W,2020ApJ...890..117D}, and with an accompanying gas component \citep{1993A&A...268...65F,2003MNRAS.341.1179A, 2007ApJ...666..189B,2009ApJ...707..218V,2010ApJ...719.1470V,2013MNRAS.429.1949A, 2014MNRAS.438L..81A,2023ApJ...953..173B}. This exchange of angular momentum is the mechanism through which the bar slows in isolated simulations and can create a wake or shadow bar in the live dark matter halo \citep[e.g.][]{1992ApJ...400...80H, 2000ApJ...543..704D, 2002MNRAS.330...35A,2003MNRAS.341.1179A,2005MNRAS.363..991H,2016MNRAS.463.1952P,2021ApJ...915...23C}.  Many observations find bar pattern speeds are anti-correlated with observed bar length \citep{2020A&A...641A.111C,2022MNRAS.517.5660G}, while others propose that the observed trend of slower bars with longer bar length is a result of a decrease in disk scale length from fast to slow bar populations \citep{2021AJ....161..185W,2022ApJ...926...58L}. 

The evolution of central stellar bars and bulges are intertwined through their exchange of angular momentum as well as mixing their stellar populations, complicating the study of individual galaxy's histories \citep{2002ApJ...569L..83A}. The angular momentum exchange is primarily driven by the stellar bar, via resonant interactions at the Lagrange point (-1:1) and bar-associated (2:1) inner Lindblad resonance (ILR) \citep{2012MNRAS.421..333S,2013MNRAS.430.2039S}
The coevolution of stellar bars and central classical bulges has been studied in isolated simulations with static central bulges as a stabilizing component to steady against bar formation and disk destabilization \citep[e.g.][]{1973ApJ...186..467O,1992MNRAS.259..328A,2012MNRAS.421..333S}, and with live bulge components that dissipate angular momentum \citep[e.g.][]{1998MNRAS.299..499W,2001A&A...367..428A,2003MNRAS.341.1179A,2006ApJ...637..214M,2018MNRAS.477.1451F,2025MNRAS.537.1475M}. 

Observed bulge-to-total luminosities of $>20\%$ are found in $\sim1/3$ of spiral galaxies $>10^{10}M_\odot$, but less than 10\% are bulge dominated ($B/T>0.4$) \citep{2009ApJ...696..411W}. Classical bulges are a result of hierarchical formation processes and remnant of galaxy mergers. 
In $\Lambda$CDM, more than 70\% of Milky Way sized halos ($10^{12}M_\odot$) should have at least one merger event with $10^{11}M_\odot$ within the last $10$~Gyr \citep{2008ApJ...683..597S,2022ApJ...939L..31H} and 15\% have had a major merger (mass ratio 1:4) within the last $5$~Gyr \citep{2022MNRAS.516.5404S}. Once established, bulge components are generally expected to be long lived features of disk galaxies \citep{2004AAS...204.6503J,2008ApJ...675.1141S, 2011MNRAS.415.3308G}. 
Observations reveal the Milky Way has at most a classical bulge of less than 10\% of the total stellar mass \citep{1995ApJ...445..716D, 2010ApJ...720L..72S, 2012ApJ...757L...7L, 2012ApJ...756...22N, 2013MNRAS.432.2092N,2016ARA&A..54..529B,2017MNRAS.471.3988C,2021MNRAS.501.5981L} and up to 70\% of massive local disk galaxies completely lack a classical bulge \citep{2004ARA&A..42..603K, 2009PASP..121.1297K, 2010ApJ...723...54K, 2010ApJ...720L..72S, 2011MNRAS.415.3308G}. The low fraction of observed classical bulges in local disk galaxies challenges our understanding of hierarchal galaxy formation processes \citep{2010ApJ...723...54K,2022ApJ...939L..31H}.

In this work, we use a suite of high-resolution N-body simulations to quantify the result of bar-driven angular momentum transfer to classical bulges under varying initial conditions. 
Our results provide timescales for the bar-bulge co-evolution processes and constrain the longevity of central, classical stellar bulges in the presence of strong-stellar bars, especially in the local secular evolution dominated universe. We also present these results in the context of local galaxy integral field unit (IFU) observations of barred galaxies. Our findings have implications for understanding the morphological evolution of disk galaxies, the formation history of the Milky Way, and the broader context of galaxy evolution.

The remainder of this paper is organized as follows: Section 2 describes our simulation setup and parameter choices; Section 3 presents our results on bulge transformation efficiency; Section 4 discusses the implications for observations and theory; and Section 5 summarizes our conclusions.

\section{Methods} 
    \subsection{Simulation Parameters}\label{sec:sim params}
\begin{figure*}
        \centering
        \includegraphics[width=1\textwidth]{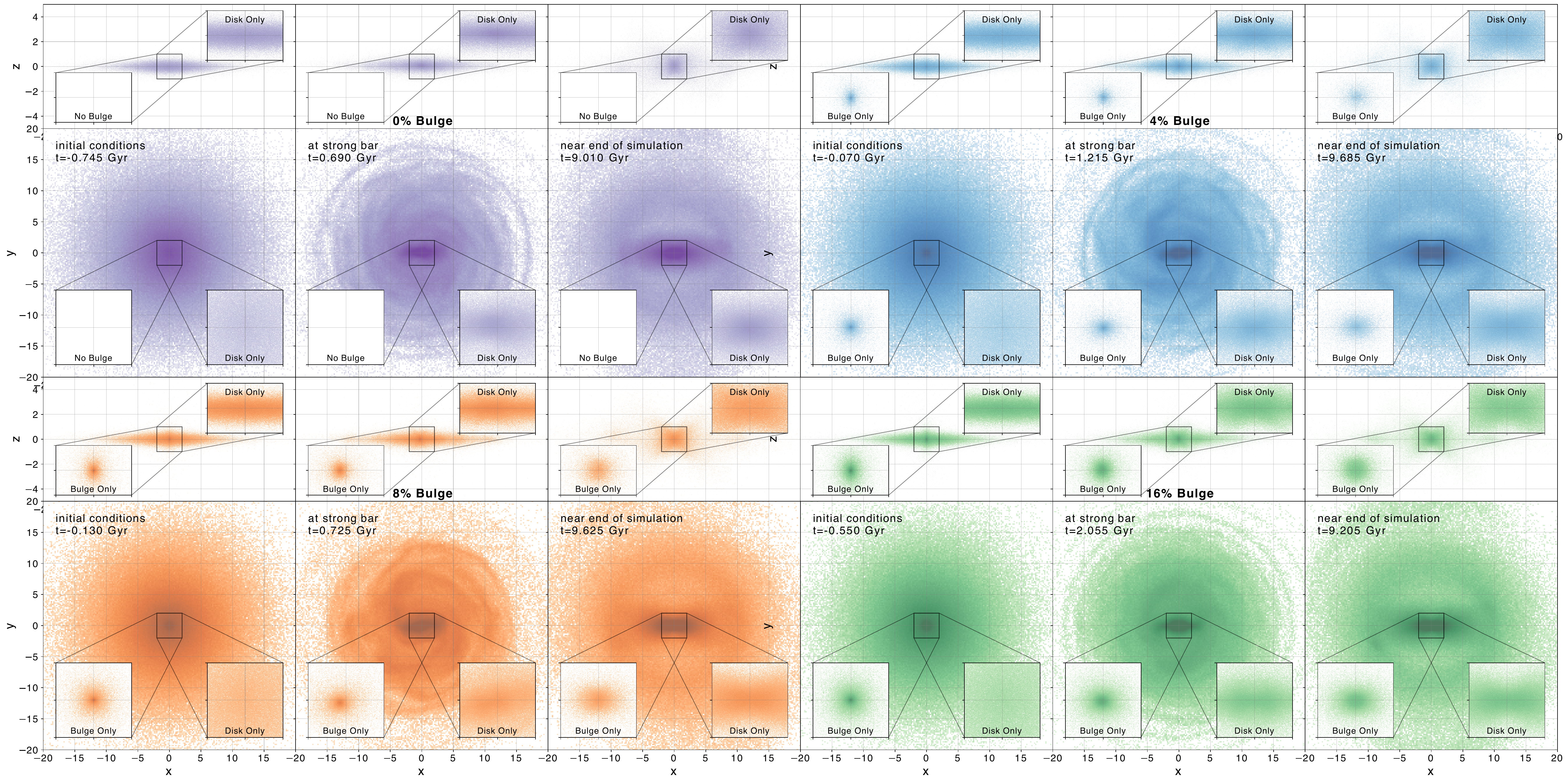}
        \caption{Each panel shows the overall density distribution of each model at three times: the initial conditions, the time of each bar reaches the strong-bar threshold of $A_{\rm bar}=0.2$, and $250$~Myr before the end of each simulation. The models have the same disk mass but vary in the bulge mass fractions ranging from 0\% Bulge (\textbf{purple}, top left), 4\% Bulge (\textbf{blue}, top right), 8\% Bulge (\textbf{orange}, bottom left), 16\% Bulge (\textbf{green}, bottom right). Each sub-panel of the plots has inset frames of the bulge and disk stellar populations, while the background frame shows the combined stellar populations. All frames use the same fixed color saturation with logarithmic scale to emphasize non-axisymmetric features. Each model develops a strong bar and a substantial \bpx{} pseudobulge visible in the edge-on top panels. The central component of the stellar bar is dominated by the bulge in models where it is included; the 0\% Bulge model evolves a central concentration with similar bulge-like structure to the models with included central classical bulges.}
        \label{fig:sims}
    \end{figure*}

    We simulate four isolated disk galaxies, each with a three component setup of live-dark matter halo, stellar disk, and central stellar bulge of varying mass fraction, including one model with no central initial bulge. Each set of initial conditions was created with a modified version of MakeNewDisk \citep{2005MNRAS.361..776S} and then subsequently evolved with \textsc{arepo} \citep{2010MNRAS.401..791S,2016MNRAS.455.1134P,2020ApJS..248...32W}. 
    Each disk is a radially exponential and vertically isothermal stellar disk of mass $4.8\times10^{10}\,M_{\odot}$, radial scale length of $\sim2.7\,\textrm{kpc}$, and vertical scale height of $\sim0.32\,\textrm{kpc}$. The halo and bulge are both \citet{1990ApJ...356..359H} profiles: the dark matter halo has mass $10^{12}\,M_\odot$, $R_{200}=163\,\textrm{kpc}$ and concentration parameter $c=11$, and each bulge has a scale length $3.15\,\textrm{kpc}$. The bulge masses are $0$, $2$, $4$, and $8\times10^9\,M_{\odot}$ corresponding to $0\%$, $4\%$, $8\%$, and $16\%$ of the disk mass, respectively. The initial conditions of each simulation are shown in purple, blue, orange, and green, respectively, in the left-most frames of each set in Figure~\ref{fig:sims}. For further detailed description of the simulations and consistency checks, we refer the reader to \cite{2025MNRAS.537.1475M}, with analysis on these same simulation models, as well as \cite{2020ApJ...890..117D} and \cite{2023ApJ...953..173B}.

    In each simulation that includes a bulge, to isolate the effect of secular evolution after bar formation we performed a controlled experiment: 
    at the snapshot where the bar first reaches $A_{\rm bar} = 0.2$ (see below), we reinitialized the simulation by replacing bulge particles with their original initial condition counterparts, ensuring radially isotropic  orbits. The results of this test are discussed in Section \ref{sec:bulge bar evolution}. 

    \subsection{Stellar Bar Properties}
    Each model develops a strong stellar bar within $3$~Gyr of the simulation start. We measure the bar amplitude, $A_{\rm bar}$, in cylindrical radial rings of width $250$~pc as the amplitude of the $m=2$ mode of the stellar surface density, $A_2$, each disk normalized by the average surface density, $A_0$, 
    \begin{equation}
        A_{\text{\rm bar}} = \frac{1}{N} \Bigg | \sum_j e^{2 i \phi_j} \Bigg |,
        \label{eq:A2}
    \end{equation}
    over the $j^{th}$ stellar disk particle \citep{1986MNRAS.221..195S, 2006ApJ...645..209D}. 

    We define the bar length, $R_{\rm bar}$, as the radius where $A_{\rm bar}$ drops to $20\%$ of its peak value. Early in the simulation, the $m=2$ spiral arm modes have stronger amplitudes relative to the central bar such that a 20\% threshold leads to an over-estimate when these spirals align with the central bar. This technique provides an analogous bar-length measurement to one used in Milky Way observations \citep[e.g.][]{2015MNRAS.450.4050W}.%
    We utilize two times to compare across the individual simulations: the bar-start (defined as $t=0$~Gyr) as the time after which $A_{bar}$ is generally increasing, and the time the bar breaches the strong-bar threshold of $A_{\text{bar}}=0.2$ \citep{2001A&A...367..428A, 2023ApJ...947...80B}. 

\begin{figure*}
        \centering
        \includegraphics[width=1\linewidth]{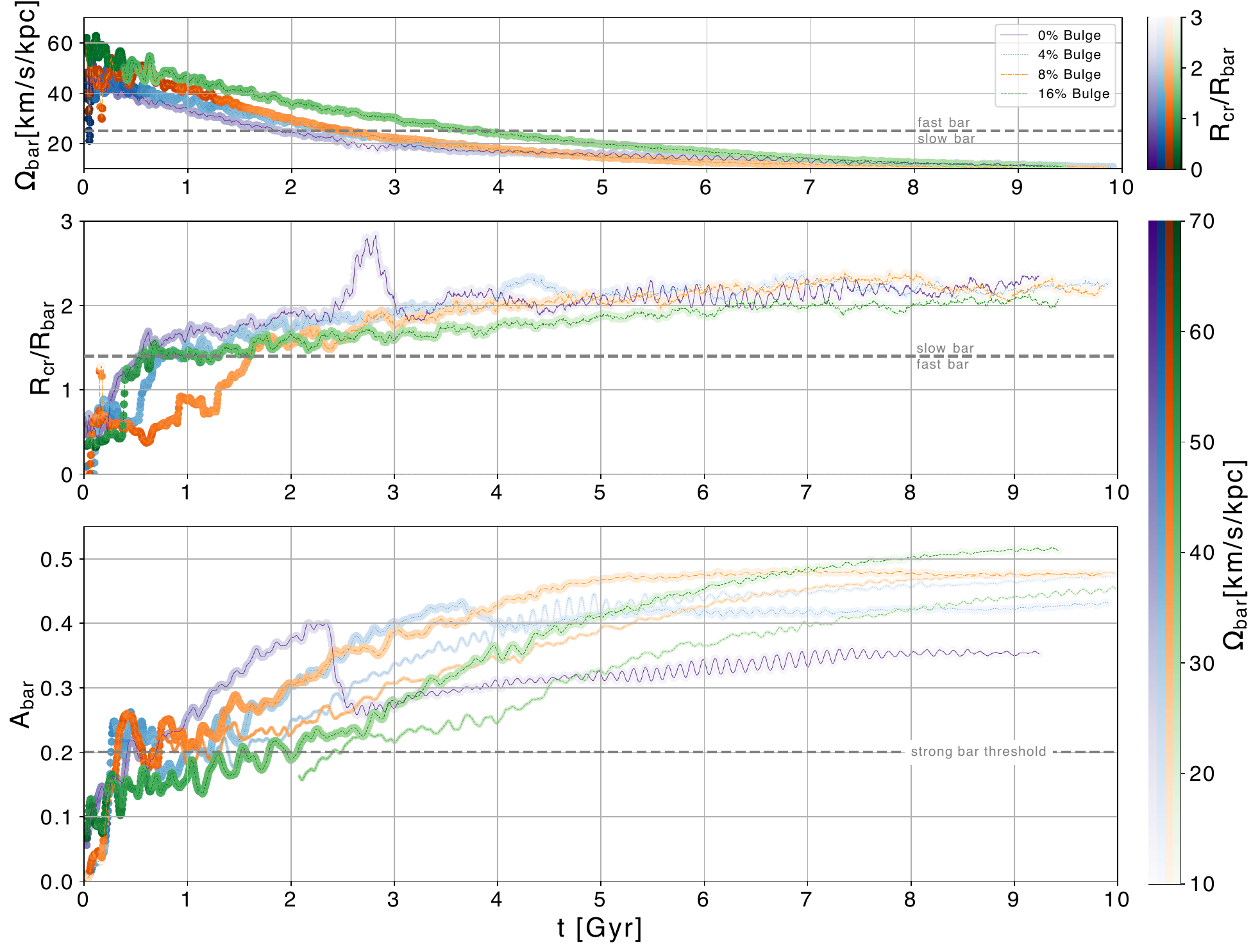}
        \caption{The bar pattern speed ($\Omega_{\rm bar}$), ratio of the bar corotation and radius ($\mathcal{R}$), and the bar amplitude ($A_{\rm bar}$) are shown over time since bar-start, the time after which the bar is growing. The models shown include 0\% Bulge (\textbf{purple, solid} line), 4\% Bulge (\textbf{blue, dotted} line), 8\% Bulge (\textbf{orange, dashed-dot} line), to 16\% Bulge (\textbf{green, dashed} line). Matching markers along the lines are colored with ranging intensity for \textit{fast} to \textit{slow} bar parameters, ranging dark to light. The top panel color corresponds to the $\mathcal{R}$ measurement and includes a horizontal dashed gray demarcation line at the $\Omega_{\rm bar}=25$~km/s/kpc. The middle and lower panel lines are accompanied by marker colors corresponding to the bar pattern speed ($\Omega_{\rm bar}$) with a horizontal dashed gray demarcation line at the $\mathcal{R}=1.4$ threshold. The lower panel also includes a second set of thin lines with matching line styles for the re-initialized 4\%, 8\%, and 16\% Bulge models diverging from the original model lines; this panel also features a horizontal dashed gray demarcation line at $A_{\rm bar}=0.2$.}
        \label{fig:Abar}
    \end{figure*}

    The bar pattern speed, $\Omega_{\rm bar}$, is computed from the time derivative of the phase of the $m=2$ mode at $0.5$~Myr intervals. The pattern speeds decrease over time, as shown in the top panel of Figure~\ref{fig:Abar}, with each starting in the fast-bar regime of $\Omega_{\rm bar}>25$~km/s/kpc and reaching the slow-bar regime within $3$~Gyr. 
    
    We also classify each stellar bar at all times through the ratio of the corotation radius of the bar and the bar length,
    \begin{equation}
        \mathcal{R} = R_{\rm CR}/R_{\rm bar},
        \label{eq:curlyR}
    \end{equation}
    shown in middle panel of Figure~\ref{fig:Abar}. The corotation radius, $R_{\rm CR}$, is taken as the radial location where the the circular velocity from the total gravitational potential is equal to the bar pattern speed. Each model crosses the observational threshold for slow-bars ($\mathcal{R}=1.4$, \citealt{2000ApJ...543..704D}) within 1.6 Gyr of bar formation. 
    We include both metrics of the bar rotation speed but emphasize that $\Omega_{bar}$ is a more direct comparison between our models and the included observations.

    Of our four models, all evolve a significant Boxy-Peanut-X-Feature (\bpx) pseudobulge component. The No-bulge and 4\% Bulge models exhibit classical bar-buckling events, while the 8\% and 16\% do not. Models with higher mass initial bulges exhibit greater stabilization of the bar-evolution and promote the formation of a pseudobulge through the passage from the horizontal (h2:1) resonance and through the vertical (v2:1) resonance \citep{2025MNRAS.537.1475M}.  

    \subsection{Orbital Resonance Frequency Populations}
    The stellar bar is populated by stars on stellar orbits that are both in resonance with the bar, and inherently a part of the bar itself, with ratios of two radial maxima per azimuthal period, $2\Omega_r=\Omega_x$ as measured in the fixed-rotating frame of the bar.
    
    We characterize individual stellar orbits throughout each model in time-windows set dynamically by the bar period as $1.5\times2\pi/\Omega_{\rm bar}$ for the central time in each window. Our analysis captures the evolving resonances in the stellar orbits by measuring the fundamental frequencies for each component in each star's orbital components: $\Omega_r$, $\Omega_z$, and $\Omega_x$ as computed with respect to the x-axis in the bar-frame (with the bar angle subtracted off each star's azimuthal coordinate to align it with y=0). We utilize \texttt{naif} frequency analysis code \citep{2023ApJ...955...38B} to extract these frequencies, for more detailed discussion and Figures showcasing the morphology of orbital families of interest in these simulations, we direct readers to \cite{2025MNRAS.537.1475M}. 

    We perform this frequency characterization on the populations initialized as disk and bulge \citep[for details on initialization of disk and bulge see][]{2005MNRAS.361..776S} as they evolve in each model in our simulation suite, including the reinitialized bulge simulation models. The population fraction of the $2\Omega_r=\Omega_x$ stellar orbits of the disk increases at the same time and rate as the growth of the bar amplitude $A_{\rm bar}$, demonstrating that this frequency selection is successful at capturing the bar-supporting orbital population. 

    \subsection{Differentiating Initial Bulge from Disk Populations}\label{sec:PDiskBulge assigned}
    The initial conditions of our simulation include separate bulge and disk star particles. As the galaxy evolves, 
    early disk and classical bulge stellar populations will radially mix and gravitationally interact. 
    
    In this work, we estimate the probability that each star would be identified as bulge or disk using a Bayesian classifier.
    We use 
    the $r$, $z$, $v_z$, and $v_\phi$ 
    coordinates of each stellar particle to determine the likelihood function as a kernel density estimator for each point of the initial conditions of the bulge population, $P(r,z,v_z,v_\phi|\rm Bulge_{init})$, and at each time step of the disk component, $P(r,z,v_z,v_\phi|\rm Disk_{t})$. We input these conditional probabilities into Bayes' theorem with their relative population fractions as the prior for each population probability. From this we recover a posterior probability for each of the two populations, given any set of $r$, $z$, $v_z$, and $v_\phi$ phase space coordinates. 
    This assigns a probability to any combination of the coordinates of a stellar particle at any time that it is disk star, $P(\rm Disk)$, or a bulge star, $P(\rm Bulge)$.
    We compare the distributions of $P(\rm Disk)$ for the bar-resonance stellar population (h2:1) to provide a metric of how ``bulge-like'' and ``disk-like'' these resonant populations would appear in an observational sample of their kinematic components. 

    \subsection{Observational Barred Galaxy Selection}
    \label{sec:manga}



    To verify whether we see evidence of secular attrition of classical bulges happening in observations, we use the sample of barred galaxies of \cite{2023MNRAS.521.1775G}, identified in Galaxy Zoo DESI (GZ DESI; \citealp{walmsley_2023}). Bar kinematics were measured using the Tremaine-Weinberg (TW) method \citep{1984ApJ...282L...5T} applied to the Sloan Digital Sky Survey (SDSS-IV; \citealt{blanton_2017}) Mapping Nearby Galaxies at Apache Point Observatory (MaNGA; \citealp{bundy_2015}) integral-field spectroscopic survey.
    
    After calculating the pattern speed, $\Omega_{\rm bar}$, \citet{2023MNRAS.521.1775G} also measured the corotation radius ($R_{\rm CR}$). 
    This was found by first fitting a two-parameter arctan model for the rotation curve of the galaxy \citep{courteau_1997}, and then finding the radius at which the angular velocity equals the bar pattern speed.

    \citet{2023MNRAS.521.1775G} also measure the dimensionless parameter $\mathcal{R} = R_{\rm CR} / R_{\rm bar}$, where $R_{\rm CR}$ is the corotation radius and $R_{\rm bar}$ is the deprojected bar radius. As discussed above, this metric is often used to delineate slow ($\mathcal{R} > 1.4$), fast ($1.0 < \mathcal{R} < 1.4$) and ultra-fast ($\mathcal{R} < 1.0$) bars (e.g. see \citealp{2000ApJ...543..704D, rautiainen_2008, aguerri_2015}). Due to the small number of ultra-fast bars in our sample, we group them with the fast bars for the remainder of this work. The final sample used in this work consists of 210 galaxies with reliable measurements of $\mathcal{R}$. For a more detailed description of the calculation of $\Omega_{\rm bar}$, $R_{\rm CR}$, and $\mathcal{R}$ in this sample, we refer the reader to \cite{2023MNRAS.521.1775G}.

    These 210 galaxies all have clearly resolved bars, as well as bulges of various sizes. One of the questions asked in the decision tree of Galaxy Zoo 2 (GZ2; \citealt{willett_2013}) is whether there is a central bulge in the galaxy and, if so, how large it is. These vote fractions were used in \cite{masters_2019} to define a bulge prominence parameter. However, the Galaxy Zoo decision tree changed slightly between GZ2 and GZ DESI to take advantage of the greater depth that the DESI survey provides, compared to SDSS. Volunteers could now vote on whether the bulge was small, moderate, large, dominant, or no bulge at all. Due to differences in survey depth, we adopt the revised definition of bulge prominence from Garland et al. (in prep), based on updated Galaxy Zoo DESI vote fractions.

    \begin{equation}
        B = 0.2\;p_{ \rm small} + 0.5\;p_{\rm moderate} + 0.8\;p_{\rm large} + 1.0\;p_{\rm dominant}\:,
        \label{eq:bulge_eq}
    \end{equation}
    where $p_{ \rm small}$, $p_{ \rm moderate}$, $p_{ \rm large}$, $p_{ \rm dominant}$ represent the fraction of people that voted that the bulge was small, moderate, large, or dominant, respectively. We use these measurements of $\mathcal{R}$, $\Omega_{\rm bar}$, and bulge prominence of these 210 galaxies in Section \ref{sec:slow bars small bulges} to determine whether evidence of bulge attrition is found in the observations. Note that there was not a single galaxy out of the 210 galaxies in our sample where more than half of the volunteers said that there was no bulge (i.e. $p_{\rm no\;bulge} > 0.5$), while only 4 had $p_{\rm no\;bulge} > 0.1$.
    
\section{Results}\label{sec:results}
We present results quantifying the coevolution of stellar bars and classical bulges, focusing on angular momentum exchange and the trapping of bulge stars into bar-supporting resonant orbits. 
We also assess the observability of this transformation and its kinematic distinction from pseudobulge populations.

\subsection{Co-Evolution of the Central Bulge and Stellar Bar}\label{sec:bulge bar evolution}
Our suite of N-body models are initialized with a bulge component of 0\%, 4\%, 8\%, and 16\% of the disk mass. By the end of the evolution, we find that for each model with an initial bulge, the central bulge is an integral part of the stellar bar by the time the bar reaches the strong bar threshold. This is visible in Figure \ref{fig:sims} as the inset plots show the initial bulge and disk populations for each model at each included time. The 0\% Bulge model that does not include a classical bulge forms an apparent central bulge component by the time of the strong bar threshold and exhibits a more centrally concentrated apparent bulge than each of the models that include an initial classical bulge. The initial bulge forms the central component of the stellar bars in each of the models where it is included. 

    \begin{figure*}
         \centering
         \includegraphics[width=1\textwidth]{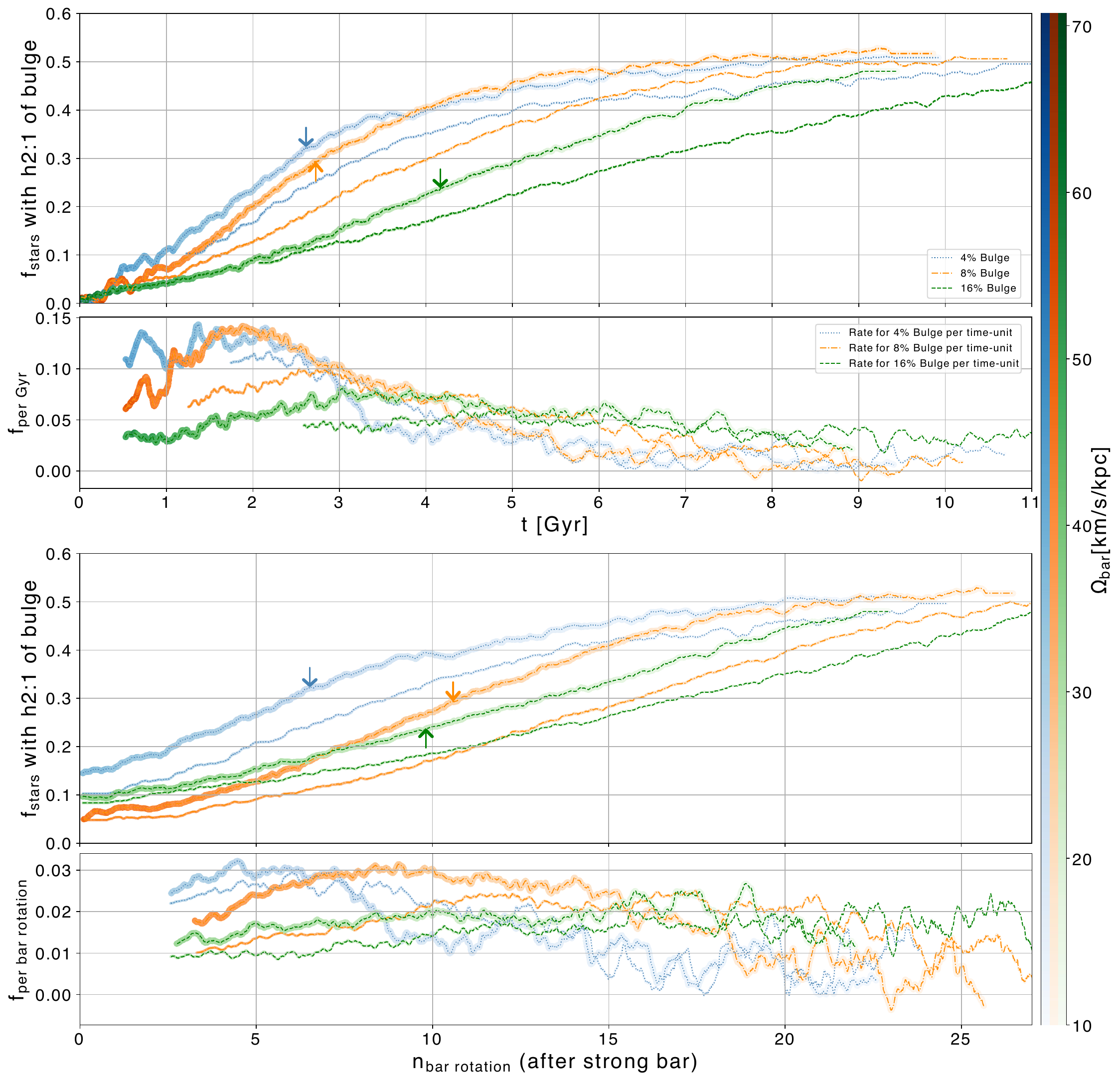}
            \caption{For the 4\%, 8\%, and 16\% Bulge models, the fraction of stars with h2:1 bar-supporting stellar orbits is shown over time (top panels) and as a function of the number of bar rotations, after the time the strong bar threshold is reached (bottom panels). Each set includes a panel showing the rate of attrition as the rate of increase in the fraction of the bulge that is h2:1, taken over $1$~Gyr time differences. Each line is accompanied by colored markers with increasing saturation corresponding to faster bar pattern speed, small arrows mark the time each model bar crosses the $\Omega_{\rm bar}=25$~km/s/kpc threshold from fast to slow bars. Both panels feature thin secondary tracks for the reinitialized bulge model and horizontal gray dashed reference lines of 50\% of bulge population. All models converge to 50\% of each bulge exhibiting h2:1 bar-supporting orbital frequencies. The rate of increase in bar-supporting orbits within the initial bulge is initially highest in the models with lowest bulge fraction ($\sim15\%$ per Gyr) and decreases in the 4\% and 8\% Bulge models as the bar slows over the first $5$~Gyr; the rate is approximately constant ($\sim5\%$ per Gyr) in the 16\% Bulge model as the more significant Bulge steadies the rate of capture. The attrition rate measured over bar-periods (lower panel) is similar across models ($<3\%$ per bar period) and steadier than the rate as measured over time. The initial rate of attrition is highest for the models with smallest mass bulge components and is varies the least in the 16\% Bulge model.} 
            \label{fig:bulge_h21}
    \end{figure*}

An initially isotropic, non-rotating, classical bulge has a symmetric distribution of azimuthal velocities centered around zero. Over many orbital interactions with the accompanying disk, the bulge absorbs some of the rotational velocity. The angular momentum exchange is primarily driven by the stellar bar, via resonant interactions at the Lagrange point (-1:1) and bar-associated (2:1) inner Lindblad resonance (ILR) \citep{2012MNRAS.421..333S}. In this work we use the notation h2:1 to denote the horizontal, in-plane, ILR and v2:1 to denote the analogous vertical resonances of the bar. 

We find that up to 50\% of the initial classical bulge stellar population becomes trapped on orbits satisfying the h2:1 resonance condition, the fraction of each model's initial bulge population that exhibits h2:1 is shown in Figure \ref{fig:bulge_h21}. The rate of trapping is lowest in the 16\% Bulge model; the 8\% and 4\% Bulge models each have initial rates of $10\%-15\%$ per Gyr. The trapping is efficient enough early in the evolution that within 3~Gyr of crossing the strong-bar threshold, the h2:1 fraction rises from 10\% to 30\%. The efficiency of the trapping is more similar across models when measured per bar period (see lower panels each set in Figure \ref{fig:bulge_h21}), though still higher in the presence of less massive bulge components. The 16\% Bulge model has a steady trapping rate throughout the evolution of around 5\% per Gyr and 2\% per bar period. The massive bulge moderates the trapping process, just as it moderates the bar evolution and growth of the pseudobulge. The less massive the initial classical bulge, the more rapid the attrition of the bulge by the bar due to the more efficient bar evolution in the presence of less significant bulge components. 

For each model, we repeat our analysis with the reinitialized bulge from the time each bar reaches the strong bar threshold to test if the trapping is tied to the initial formation of the stellar bar within the disk or if this process occurs throughout the bar evolution. These additional models each have 50\% of the bulge component exhibiting h2:1 frequencies on the same timescales as the original simulations, as shown in the included thin lines of each panel of Figure \ref{fig:bulge_h21}. The rate of trapping for the reinitialized models is similar across all models as measured over bar periods and is only slightly lower than in the original models. The attrition rate in the reinitialized models does not have as sharp of an increase in the 4\% and 8\% Bulge models as in the original models but the rates still reach above 10\% per Gyr with these reinitialized bulge components. 

 \begin{figure}
     \centering
     \includegraphics[width=1\linewidth]{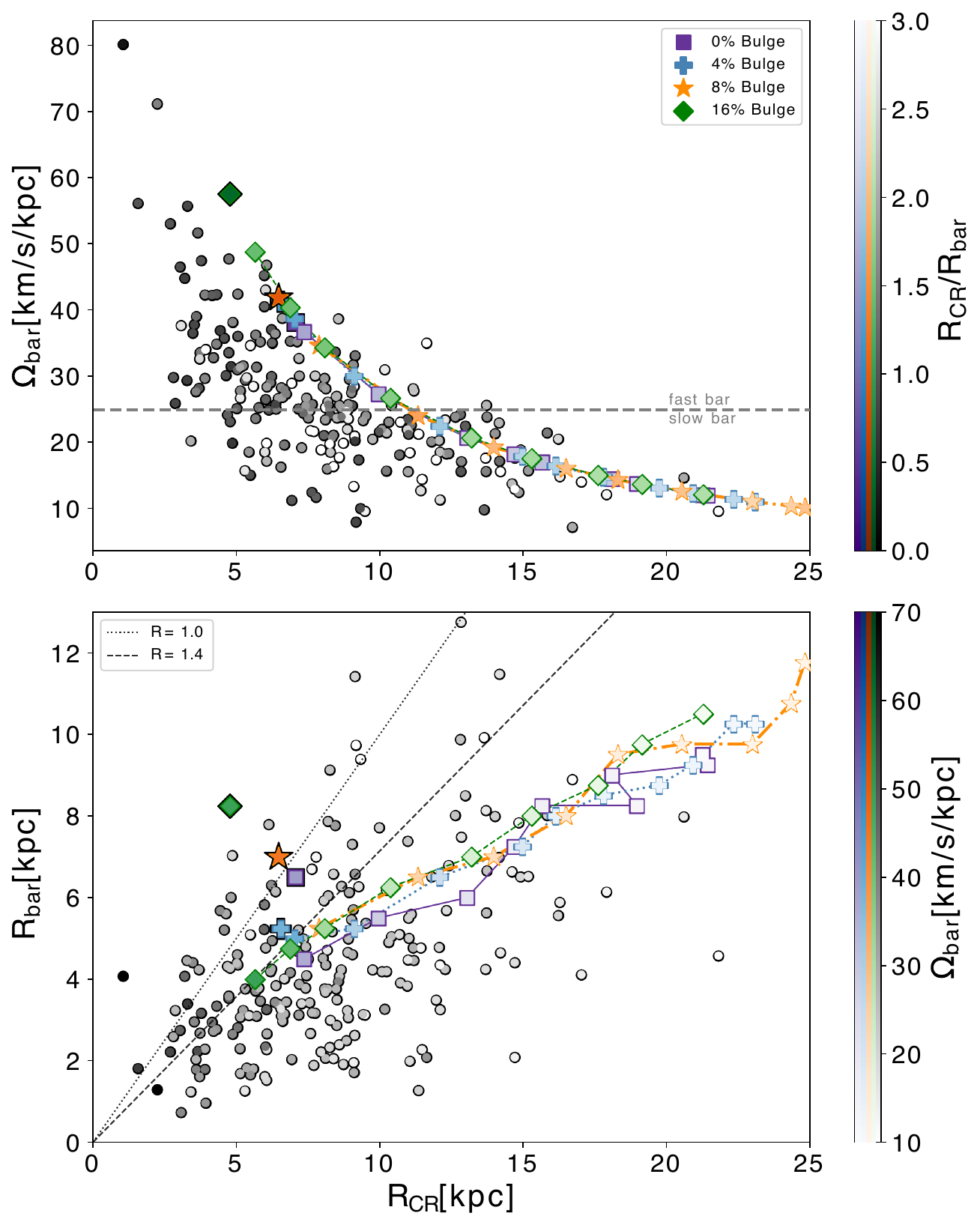}
     \caption{The bar pattern speed, $\Omega_{\rm bar}$, and bar radius, $R_{\rm bar}$, are shown as a function of the corotation radius, $R_{\rm CR}$, in the top and bottom panel, respectively. The observational sample of barred galaxies of \cite{2023MNRAS.521.1775G} is shown in the background in each figure as round data points with increasing saturation corresponding to faster bar speeds as $\mathcal{R}$ in the top panel and $\Omega_{\rm bar}$ in the lower panel. The $\mathcal{R}$ is the ratio of $R_{\rm CR}$ and $R_{\rm bar}$ so the threshold values of 1.4 (fast-bar) and 1.0 (ultra-fast bar) are marked with dashed and dotted lines, respectively.  Each of our four models are shown with one point in the time when the $m=2$ bar-mode is detectable but before the bar is steady in its evolution; these fall in the ultra-fast or fast regime and are outlined in black. Each simulation is then plotted with tracks from the time when the bar radius is smallest that is followed by growth, and then at $1$~Gyr time-steps following with markers of matching scaled saturation: 0\% Bulge (\textbf{purple, square}), 4\% Bulge (\textbf{blue, plus}), 8\% Bulge (\textbf{orange, star}), and 16\% Bulge (\textbf{green, diamond}). The simulation models occupy a narrow range of the observational data due to their shared initial conditions (except the included Bulge mass) but otherwise follow the same trends as the observational sample.}
     \label{fig:fast_slow_wobsv}
 \end{figure}

With higher initial bulge fraction in our models, the bars transition from fast ($\Omega_{\rm bar}>25$~km/s/kpc) to slow ($\Omega_{\rm bar}<25$~km/s/kpc) pattern speeds at later times. Each bar is classified as slow through the ratio of the corotation radius and bar length ($\mathcal{R}>1.4$) much earlier in their evolution than when measuring the pattern speed directly. We show the evolution of bar pattern speed and $\mathcal{R}$ for each model in Figure \ref{fig:Abar}. The values of pattern speed decrease from $60$ to $10$~km/s/kpc while $\mathcal{R}$ increases from ultra-fast ($\approx 0.5$) to the bar-slow regime  ($\mathcal{R}>1.4$) within $1$--$2$~Gyr. The early classifications of ultra-fast may be contaminated by biases from the early, strong, $m=2$ spiral structure, as suggested by \cite{2020MNRAS.497..933H} 

 Figure \ref{fig:fast_slow_wobsv} shows the evolution of 
 $\mathcal{R}$ and $\Omega_{\rm bar}$ across models in the top and bottom panels, respectively, overlaid on the observational distributions from the barred galaxy sample.  
 We present the bar speed values for each model from the time the bar is at a minimum radius after the bar-start, and at $1$~Gyr increments following this. Each model stabilizes in the range $1.4<\mathcal{R}<2.3$ within $1$~Gyr. Before the bar is steadily evolving, the measured length is augmented by the emergence of $m=2$ spiral arm modes attached to the central nascent stellar bar; for discussion of this bias in observations see\cite{2020MNRAS.497..933H}. Due to these biases, the early bar period is classified in the ultra-fast ($\mathcal{R}<1.0$) or fast ($\mathcal{R}<1.4$) regime; we include one point in this regime for each model as well in both panels for completeness. 
 
 Regardless of bulge fraction, each early bar initially is rapidly rotating at $\gtrsim40$~km/s/kpc. The bars slow most efficiently in the lower bulge fraction models, reaching $<25$~km/s/kpc within $3$~Gyr. The most significant bulge (16\%) exhibits slower bar evolution and slows at a later time. Lower initial bulge masses correlate with more rapid angular momentum transfer and more efficient secular attrition.%

\subsection{Slow Bars are Accompanied by Small Bulges}\label{sec:slow bars small bulges}

\begin{figure}
    \centering
    \includegraphics[width=\linewidth]{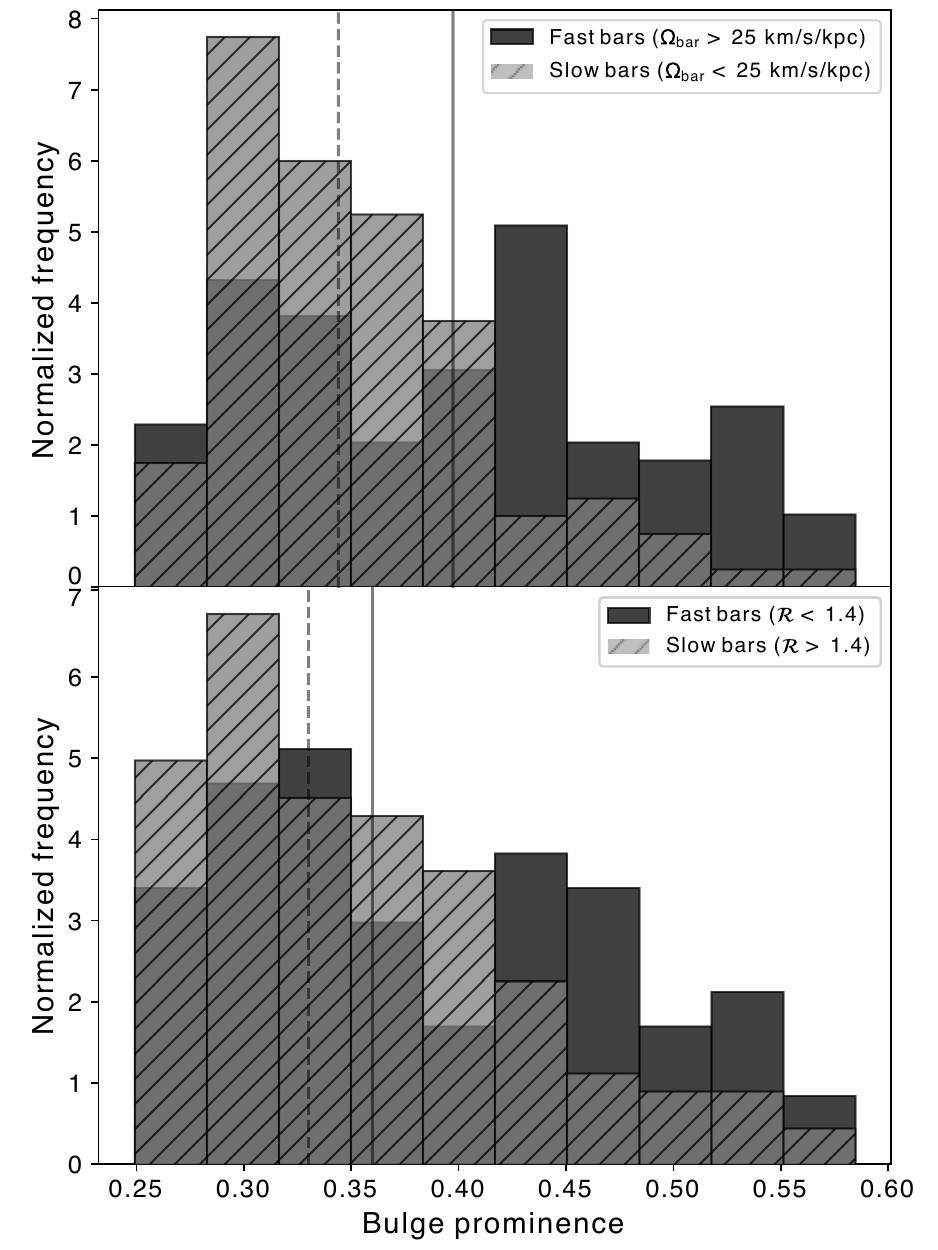}
    \caption{The top panel shows the bulge prominence ($B$) as defined in Equation \ref{eq:bulge_eq} of the fast bars ($\Omega_{\rm bar} > 25$ km/s/kpc; \textbf{dark-gray}) and slow bars ($\Omega_{\rm bar} < 25$ km/s/kpc; \textbf{light-gray with slash hatching}). The dashed vertical lines indicate the median of the two populations. The bottom panel shows is similar, but for fast and slow based on $\mathcal{R}$ instead. $\Omega_{\rm bar}$ and $\mathcal{R}$ were measured in \cite{2023MNRAS.521.1775G}. An Anderson-Darling test between the two distributions in the bottom panel is $2.9\sigma$, and $>3.2\sigma$ for the top panel.}
    \label{fig:fast_slow}
\end{figure}

Slow bars are indicative  of older systems as bar slowing occurs through cumulative angular momentum exchange over Gyr timescales \citep{1972MNRAS.157....1L, 2003MNRAS.341.1179A, 2006ApJ...637..214M, 2013MNRAS.429.1949A, 2017MNRAS.469.1054A}. 
We use the MaNGA sample, described in Section \ref{sec:manga}, to observationally verify whether slower bars - which are likely older bars - have smaller bulges. This is done in Figure \ref{fig:fast_slow}, where we compare the bulge prominence, as described in Equation \ref{eq:bulge_eq}, for fast and slow bars. 

We separate bars into fast and slow based on $\mathcal{R}$ (lower panel of Figure  \ref{fig:fast_slow}) and based on $\Omega_{\rm bar}$ (top panel of Figure \ref{fig:fast_slow}). We find that the median bulge prominence of the slow bar population is lower than that of the fast bar population, regardless of whether we used $\Omega_{\rm bar}$ or $\mathcal{R}$. However, after using an Anderson-Darling 2-Sample test, we find that only the distributions divided by $\Omega_{\rm bar}$ are statistically different ($>3.2\sigma$), while the distributions divided by $\mathcal{R}$ have a statistical significance of $2.9\sigma$. 

 \begin{figure}
     \centering
     \includegraphics[width=1\linewidth]{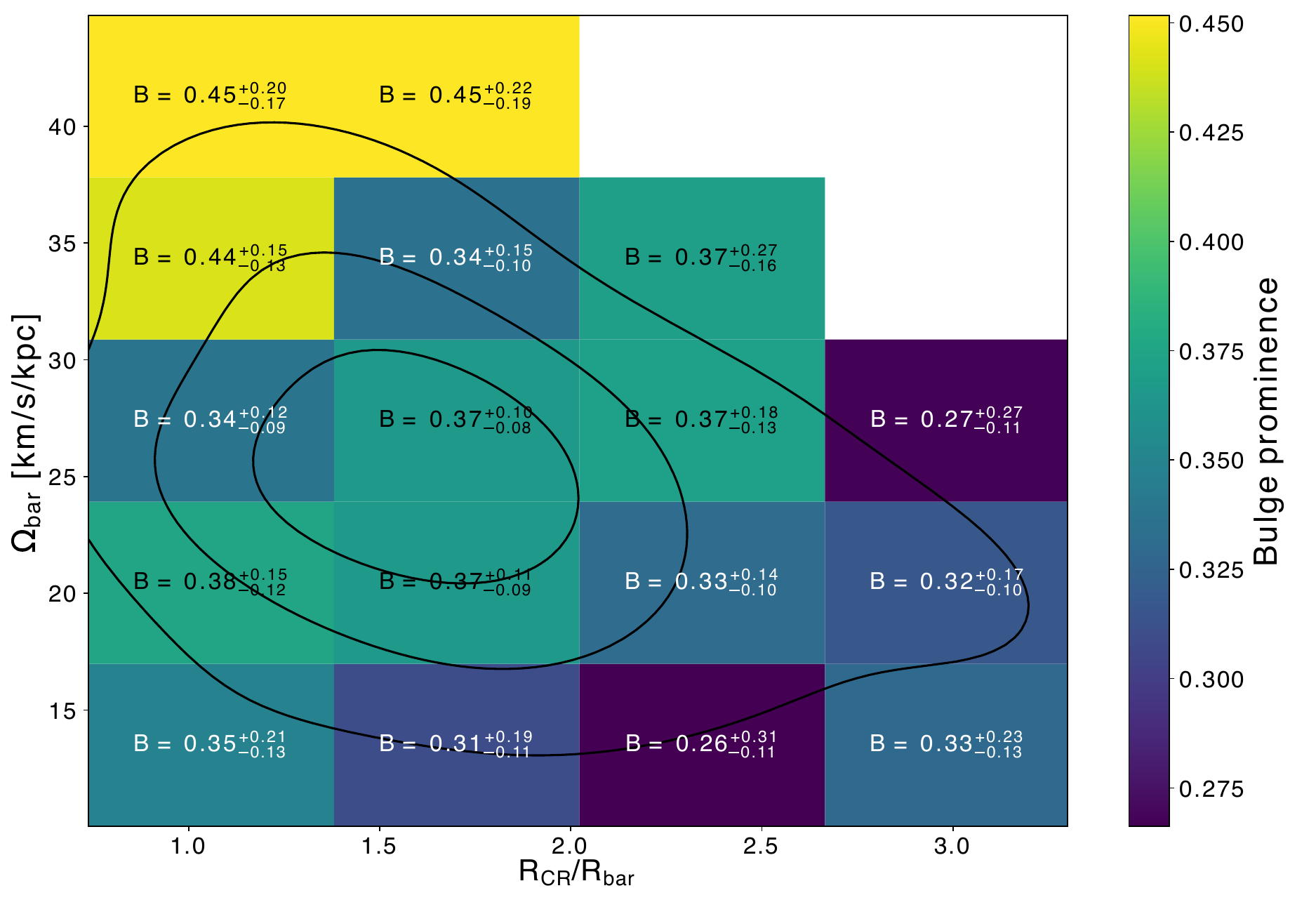}
     \caption{The mean bulge prominence is shown in bins of $\mathcal{R}$ and  $\Omega_{\rm bar}$ with uncertainty ranges reported in the Figure within each bin. The largest bulges are found among galaxies in the top-left corner of this plot, in the fast regime of each bar speed metric, while the smallest bulges are found in the bottom-right corner, in the slow regime of each metric. Each bin is colored by the bulge prominence, as defined in Equation \ref{eq:bulge_eq}. Contours are overlaid represent the quantity distribution of the data in these axis with the highest number density centered near the transition point from fast to slow bars in both metrics.}
     \label{fig:2D_plot}
\end{figure}

In Figure \ref{fig:2D_plot}, we show the median bulge prominence in different bins of $\Omega_{\rm bar}$ and $\mathcal{R}$. This also clearly visualizes that bulges tend to be larger in barred galaxies that have high values for $\Omega_{\rm bar}$ and low values for $\mathcal{R}$, i.e. fast bars. 


\subsection{Pseudo-Observational Classification of Bulge and Disk Populations}\label{sec:pseudo-observations}
\begin{figure*}
     \centering
     \includegraphics[width=1\linewidth]{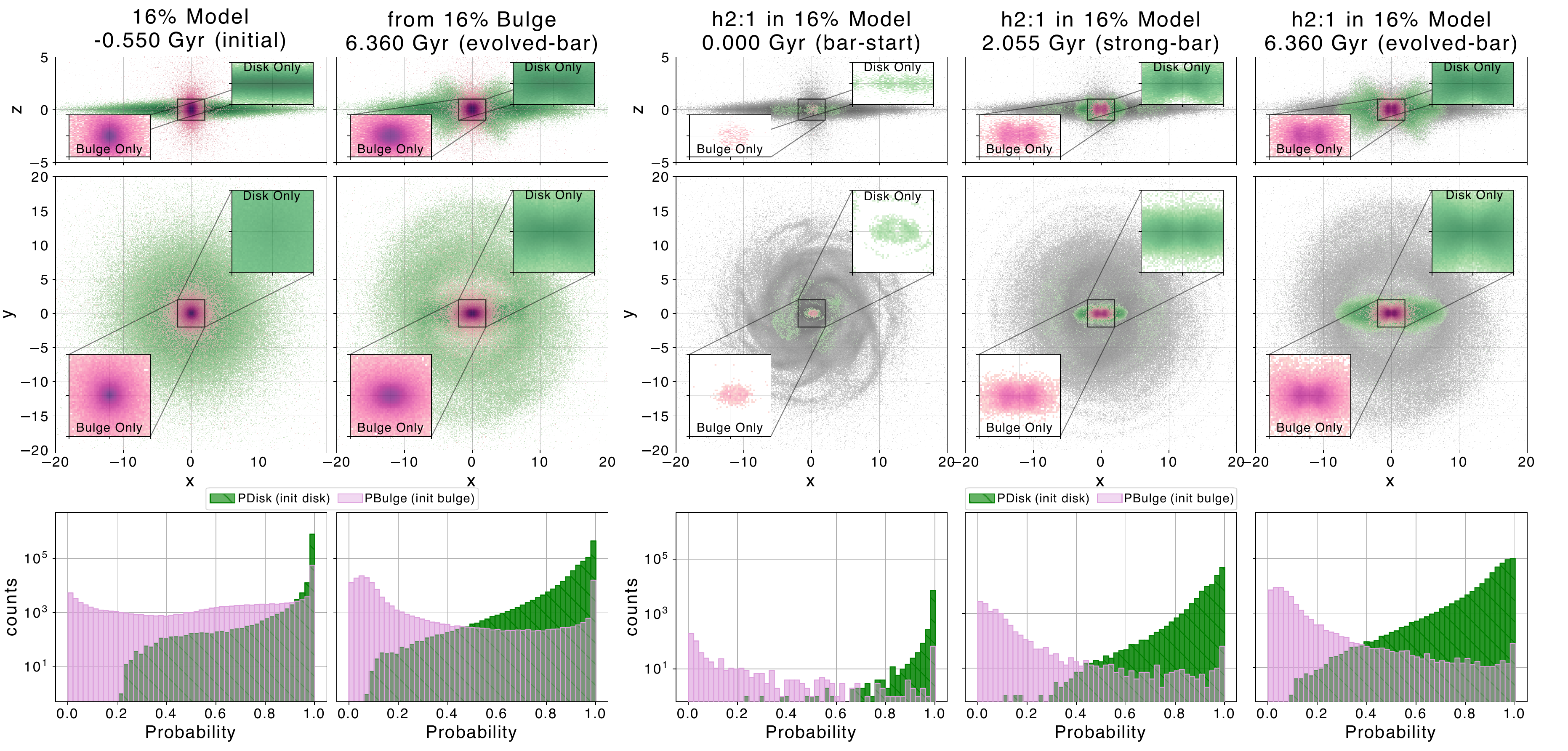}
    \caption{The initial bulge (pink) and the initial disk (green) components of the 16\% Bulge model are shown in a series of three-part panels, each of which shows the edge-on view (top row), the top-down view (middle row), and the probability that each stellar particle is identified with its initial component (bottom row). The left set of panels shows the full population of both initial stellar disk (\textbf{green}) and initial stellar bulge (\textbf{pink}) for the time of the initial conditions and the time the vertical \bpx{} pseudobulge growth is most efficient, indicating an evolved-bar. The top two panels of each include inset frames of just the initial bulge and initial disk components to highlight the morphological evolution of the disk and bulge over time. The lower panels show histogram counts and are log-scaled. They show that the disk is classified with $P(\rm Disk)>0.1$ at early and late times and that the bulge is classified with a nearly flat distribution of $P(\rm Bulge)$ with a slight increase and spike at $P(\rm Bulge)=1$ in the initial conditions frame, but over time the $P(\rm Bulge)$ develops a peak at $P(\rm Bulge)<0.2$ as the bulge component is increasingly classified as disk by our kinematic Bayesian technique. The right set of three panels show only the stellar populations that are bar-supporting (h2:1). The initial bulge stellar population dominates the central region of the bar orbits and follows the profile of the pseudobulge \bpx{} that is associated with the stellar bar at later times along with the disk population from passage of the bar.}
    \label{fig:Pfracs}
\end{figure*}

Our pseudo-observational classifier assigns probabilities to each star the star would be identified with the disk or a classical bulge population based on their $r$, $z$, $v_z$, and $v_\phi$ coordinates through Bayesian comparison with those kinematic populations in our models, as described in Section \ref{sec:PDiskBulge assigned}. We find that due to the spin-up of the bulge component, it is only very early that a majority of the initial Bulge population identified with bulge kinematics and assigned higher $P(\rm Bulge)$ than $P(\rm Disk)$. We show the distribution of probabilities from the classifier in Figure \ref{fig:Pfracs} for the initial conditions (left-most panel) with the physical distribution of the full model in edge-on (top panels) and top-down view (middle panels) with the initial Bulge and initial Disk populations in contrasting colors and featured in inset plots. The lower panel has the initial Bulge and initial Disk probability distributions with the probability that they would be successfully tagged to their initial component identity, $P(\rm Disk)$ for the initial disk component and $P(\rm Bulge)$ for the initial bulge component. Each distribution has an implicit opposite distribution, e.g. $P(\rm Disk)_{\rm initial\ bulge} =1-P(\rm Bulge)$, that we do not plot. As the bar evolves, 
the disk and bulge populations become more confused and disk stars that have been elevated into the \bpx{} region above the stellar bar are mostly classified with $P(\rm Bulge)>0.5$, as shown in the second panel of Figure \ref{fig:Pfracs}. The distribution of $P(\rm Bulge)>0.5$ orbits are dominated at all times by the stars that are initialized in the classical bulge component. By at the time each simulation has a strong-bar the fraction of the initial bulge that is kinematically distinguishable from the disk is 24\%, 28\%, and 26\% and by the end of the simulation it is 11\%, 12\% and 14\% for the 4\%, 8\%, and 16\% Bulge models, respectively.

We combine our pseudo-observational classifications with our frequency based orbital classifications to select on h2:1 bar-supporting orbits and find the distribution is dominated by $P(\rm Disk)>0.5$ and $P(\rm Disk)_{\rm initial \ bulge}>0.5$  from the time the bar is steadily evolving (bar-start) through the end of the models, shown in the right set of panels of Figure \ref{fig:Pfracs}. The initial bulge component stars with h2:1 frequencies have a probability distribution that is dominated by low $P(\rm Bulge)$-- that is, the population of bulge stars trapped in h2:1 orbits is mostly by high $P(\rm Disk)_{\rm initial\ bulge}$. The h2:1 population from the initial bulge is within the inner regions of the bar structure as seen in the top-down and edge-on perspectives of each panel while the initial disk component is smooth out to the outer bar extent. This concentration of initial bulge stars in the central region of the bar distinctly alters the light profile of the bar as the Disk Only inset shows a smooth profile from edge-on and top-down perspectives. The initial disk component that is h2:1 bar-supporting is almost exclusively classified as disk and comprises bar-orbits to the furest extent of the bar radius as the bar continues to grow and slow.

The population of bulge stars trapped in h2:1 orbits are dominated by high $P(\rm Disk)_{\rm initial\ bulge}$. The overall fraction of the initial bulge with both h2:1 and $P(\rm Bulge)<0.5$ start at less than 1.5\% in each set of initial conditions and by the end of the evolution is 22\%, 41\%, and 31\% of each initial bulge component for the 4\%, 8\%, and 16\% Bulge models, respectively. The overall fraction of the initial disk with both h2:1 and $P(\rm Disk)>0.5$ start at less than 1.6\% in each set of initial conditions and by the end of the evolution is 54\%, 59\%, and 50\% of the initial disk component for the 4\%, 8\%, and 16\% Bulge models. This trend is shown for the 16\% Bulge model from the time the bar starts to evolve in each model, at the time each measured bar is classified as strong, and at the peak time of \bpx{} evolution when the bar is significantly evolved in Figure \ref{fig:Pfracs}.

This result highlights the importance of incorporating chemical abundances when attempting to disentangle bulge and disk components observationally. With only kinematics, it is likely that the bulk of the h2:1 population of the initial bulge would be observationally categorized as disk. If the initial bulge population is chemically distinct from stellar disk, then chemical tagging could recover the bulge origin of bar-trapped stars.


\section{Discussion}
Our results demonstrate that in isolated galaxies, classical bulges can undergo significant kinematic transformation driven by the evolution of central stellar bars. This process, which we refer to as bulge attrition, involves angular momentum exchange and dynamical reshaping, not physical mass loss. As isotropic, non-rotating bulge stars become trapped in bar-supporting orbits, they acquire disk-like kinematics. This transformation alters the orbital structure and observable kinematics of the bulge, potentially leading to misclassifications in observational surveys. 

Specifically, we find that 50\% of the initial classical bulge becomes trapped in  bar-supporting orbits, specifically satisfying the h2:1 resonance condition (two radial maxima per azimuthal period). This result builds on previous work in N-body simulation models with live dark matter halos that find bars slow as they grow in length and strength in part through exchange of angular momentum with their accompanying stellar bulge and dark matter halos \citep[e.g.][]{2002MNRAS.336..785S, 2002ApJ...569L..83A, 2003MNRAS.341.1179A,2012MNRAS.421..333S,2018ApJ...858...24S,2020ApJ...890..117D}. 
As the bars in each model evolve, the fraction of the initial bulge population that remains kinematically distinguishable from the disk and pseudobulge drops to $\sim25\%$ by the time there is a strong-bar and to $\sim12\%$ within $9$~Gyr. In other words, at later times when bars are slow, only a small remnant of the classical bulge remains that retains its isotropic, dispersion-dominated character.  

We present this result within the context of observations of barred disk galaxies with bulges from the MaNGA IFU survey that reveal the slowest rotating bars are present in the galaxies with the least-significant central bulge components. In this sample, slower bars, as classified by both the ratio of corotation to bar-length and by the bar pattern speed, are accompanied by minimal classical bulges. If slower bars are indeed older, this is observational evidence for the reconfiguration of stellar bulges by evolving bars and is evidence that significant classical bulges may not be long-lived in the presence of a strong stellar bar. This result is in alignment with the low number of classical bulges in local disks compared to those with pseudobulges or no bulge at all \citep[e.g.][]{2009PASP..121.1297K,2009AN....330..100K,2010ApJ...723...54K}. This is also consistent with the very low observed classical bulge fraction in the Milky Way \citep[e.g.][]{2013MNRAS.432.2092N,2015MNRAS.450L..66P, 2017MNRAS.469.1587D,2021MNRAS.501.5981L}, especially in the context of our Galaxy's long, slow bar \citep[e.g.][]{1994MNRAS.269..753H,2013MNRAS.435.3437W,2015MNRAS.450.4050W,2020MNRAS.495..895B}.

Our models do not retain ultra-fast or fast bar classifications beyond early times, as the emerging $m=2$ mode is initially augmented by spiral arms. This supports the idea that truly ultra-fast bars may only exist briefly or under specific structural conditions not included in our limited number of isolated setups. 
This supports the theoretical proposition from \cite{2000ApJ...543..704D} that stellar bars cannot extend past $\mathcal{R}=1$, as our bars grow steadily only after they are classified in this slow regime. The bar pattern speeds therefore provide a more robust measurement of the secular evolution of our disks. Our stellar bar radii are very consistent in length across bulge fraction models, consistent with findings by \cite{2022ApJ...926...58L} that the bar length is set by the size of the host galaxy, though our bar lengths do increase as they evolve in time. Since each model in our suite shares disk-mass and scaling parameters, they converge to the same track in $R_{\rm CR}$ vs. $\Omega_{bar}$ parameter space shown in Figure \ref{fig:fast_slow_wobsv}. It may be that the disk scale-height or kinematics in our models preclude them from establishing a close-in corotation radius relative to the bar length or that only 
bars accompanied by relatively strong $m=2$ spiral arms can be classified in the ultra-fast regime or for the fast- and ultra-fast bar classifications that are common in observations.

The presence of an evolving bar significantly affects the kinematic and morphological configuration of an accompanying classical bulge on rapid timescales. We find that within $3$~Gyr, 
one third of the classical bulge stars become dynamically trapped in bar resonances. In all of our models, the fraction of the classical bulge stars with $\Omega_r=2\Omega_x$ (h2:1) increases from 5\%--10\% to 50\% over 7--9 Gyr of evolution. The more substantial the initial bulge, slower and steadier the rate of the attrition process. Substantial classical bulge components steady and slow the bar evolution process through their exchange as they delay bar evolution \citep[e.g.][]{2018ApJ...858...24S,2019MNRAS.482.1733G,2025MNRAS.537.1475M}, but once a strong bar forms, the bulge-trapping rate is similar across all models. 
We show that this process is independent of the bar-formation process as we repeat the analysis with a reinitialized, isotropic classical bulge component included after each model's stellar bar reaches the strong-bar threshold, as described in Section \ref{sec:sim params}.

Galaxy observations often conflate bar-driven secularly evolved pseudobulges with random motion supported classical bulges. Our results further emphasize the need for chemical tagging in observations of the central galactic components when possible as there is significant confusion of the classical bulge and stellar disk components in classifications that rely only on kinematic decomposition. For observations where chemical tagging will not be possible, we provide with our results updated constraints on the fraction of classical bulge populations that remain distinguishable throughout secular evolution epochs of a galaxy's history. 

In each model with a central bulge, the central portion of the stellar bar is dominated by the initial classical bulge component. We include our model that has no classical bulge component to emphasize that in the case where there is no classical bulge, the formation of a stellar bar facilitates the creation of an apparent central bulge component that is more diffuse than the central bar-bulge seen in the models with an included bulge component. The differences in resulting (pseudo)-bulge morphology and the contributions to the final bar profile are shown in Figure \ref{fig:sims} with insets of the disk and bulge for each model. The transformation of 50\% of an initial classical bulge into a rotating pseudobulge component that also contributes to the central components of the stellar bar implies that kinematic measurements of galactic bulges and classification of stellar orbits within our own Milky Way of a classical bulge component only capture a small fraction of whatever initial classical bulge component may be present early in a disk galaxy's history. While it is not surprising that there is an increase in rotation of a central classical bulge component through the angular momentum exchange with rotating stellar disk, our result can provide constraints on the history of observed bulges in galaxies throughout the local universe. Any slowed-bar accompanied by a bulge will have significantly changed the initial bulge population over the bar's evolution.
Future studies should explore the robustness of this transformation in more realistic conditions, including the presence of gas, and cosmological inflows, which could either suppress or enhance the erosion process.

\section{Conclusions}
This study quantifies the secular transformation of a classical bulge under the influence of an evolving stellar bar over long periods of secularly-dominated evolution. We do this through use of isolated N-body simulations with a live dark matter halo in which we characterize the orbital populations through resonance identification and through pseudo-observational kinematic classification of disk and bulge population probabilities. We present this work in the context of a bulge-barred observational sample in support of our key findings.

We find that classical bulges are dynamically transformed throughout the evolution of a stellar bar with up to 50\% of their stars adopting bar-like orbital characteristics. Approximately 50\% of each initial bulge in our model is classified as a bar-orbit (h2:1) and found to be classified through its kinematics as consistent with the stellar disk populations, with $P(\rm Disk)_{\rm initial\ bulge}>0.5$. These results demonstrate that an evolving stellar bar will significantly erode an accompanying classical bulge, and that any observed classical bulge component in an evolved galaxy that is accompanied by a bar may only include at most half the initially included population fraction. These simulation results are supported by observational trends from a sample of 210 MaNGA galaxies, where slow bars correlate with weaker bulge prominence. These findings are based on isolated galaxy models and reflect the outcomes of secular evolution. External interactions and the impact of gas as well as ongoing star formation --- not included here --- may alter or suppress this transformation process.

\section*{Acknowledgments}
The authors wish to thank K. Hawkins and M. Lucey for their helpful conversations in pursuit of this work. Contributions by RLM to this material are supported by the Wisconsin Space Grant Consortium under NASA Award No. 80NSSC20M0123 and the National Science Foundation Graduate Research Fellowship under Grant No. 2137424. Any opinions, findings, and conclusions or recommendations expressed in this material are those of the author(s) and do not necessarily reflect the views of the National Science Foundation. RLM also acknowledges that this work was completed on the occupied ancestral home of the Ho-Chunk and eleven other nations of the land called Teejop (day-JOPE) since time immemorial. 
KJD respectfully acknowledges that the University of Arizona is on the land and territories of Indigenous peoples. Today, Arizona is home to 22 federally recognized tribes, with Tucson being home to the O’odham and the Yaqui. 
The University strives to build sustainable relationships with sovereign Native Nations and Indigenous communities through education offerings, partnerships, and community service.
We respect and honor the ancestral caretakers of the land, from time immemorial until now, and into the future.
TG is a Canadian Rubin Fellow at the Dunlap Institute. The Dunlap Institute is funded through an endowment established by the David Dunlap family and the University of Toronto. 
Support for SL was provided by Harvard University through the Institute for Theory and Computation Fellowship.
\section*{Data Availability}
Simulations used in this analysis will be made available upon
reasonable request to the authors. The following codes were used in the analysis
and production of this manuscript: Matplotlib \citep{matplotlib}, numpy \citep{numpy}, and scipy \citep{scipy}.

\bibliography{paper}{}
\bibliographystyle{aasjournal}

\end{document}